# Retaining Earth's Habitability Beyond the Life of the Sun

**GABRIEL HARRY** 75 Chamberlain Court, 15 Ironworks Way, London, E13 9GF, United Kingdom

**Email** autonova1@yahoo.co.uk

The Earth is rare and large-scale interstellar travel may be unfeasible – instead of migrating to other solar systems, future civilisations on Earth may have to remain. However, left unchecked, the Earth's current habitability will end in one billion years. This paper investigates and conceptualises a set of megaengineering concepts required to safeguard against seven long-term threats to Earth's habitability: solar luminosity, solar engulfment, the end of tectonics, water loss, obliquity and rotation, object collision, and extra-system threats. Implementing and maintaining the set of safeguards would allow the Earth to remain habitable for 9.1 quadrillion years. This could be raised to nonillions of years via capturing interloping interstellar medium onto Jupiter. Compared with alternatives, advantages of the safeguards such as greater feasibility, lower requisite rate of energy production, and utilisation of economically viable space mining infrastructure may present remaining on Earth as a more attractive long-term survival strategy.

## 1 INTRODUCTION

The Earth has provided a cradle for life for 4.5 billion years (Gyr) - even so, the evolution of complex intelligent life on Earth may have been miraculously unlikely in such a timeframe [1], [2]. Assuming complex life survives on Earth sufficiently far into the future, any civilisation on Earth (descended from our own or not) would find itself on a planet that has exhausted at least 82% of its natural habitability time, with the Sun's increasing luminosity causing its habitable zone to leave the Earth's current orbit around 1 Gyr from now [3].

A future civilisation wishing to migrate to other stellar systems would find difficulty in finding a planet with comparable habitability to the Earth [2]. Even if one was found, it would likely be so far away that transporting a large population to it could be simply infeasible. The practicalities of transporting even a small number of people along with a supportive biosphere across multi-light year distances present enormous technological, ecological and sociological hurdles [4], [5].

Therefore, any future civilisation, or at least the majority of it, may have no option but to remain. The time that it would be able to survive on Earth would be limited by long-term threats to the planet's habitability - from the Sun, as well as additional threats. This paper investigates these threats and a set of megaengineering concepts required to collectively safeguard against them, which would dramatically increase the Earth's habitable time and circumvent the need for any future civilisation to leave the planet or solar system.

The objective is to investigate these engineering concepts, along with their requisite resources, which maximise the time that the Earth could retain its current set of attributes that make it a comfortable home for life: its surface insolation, day/night and seasonal cycle, presence of water, tectonic activity, and safety from astronomical phenomena. The engineering concepts must be possible to achieve with current theory and the resident population and biosphere of the Earth must have no requirement to leave the planet to avoid any threat.

Threats and corresponding safeguards are described in Sec. 2 for changes in the Sun's luminosity, Sec. 3 engulfment by the expanding red giant Sun, Sec. 4 the end of Earth's tectonics, Sec. 5 the loss of Earth's water, Sec. 6 changes in the Earth's axial tilt and rotation, Sec. 7 in-system object collisions, and Sec. 8 extra-system threats from explosions and encounters. The costs and benefits of the set of safeguards are then discussed, compared with the alternative strategies of migration and space habitation, as well as limitations of the study and possible avenues of future study.

## 2 LUMINOSITY

The Sun is currently 4.58 billion years old and has increased in luminosity since the start of its main sequence. In around 1 Gyr this increase in luminosity will result in the extermination of any life on Earth's surface, and will continue until the end of its main sequence in approximately 5 Gyr, at which time its hydrogen fuel will be exhausted and it will enter its red giant phase. After reaching the tip of its red giant phase, the Sun will subsequently, and relatively quickly, evolve into a white dwarf with a fraction of its present-day luminosity which would be unable to support any life at Earth's current orbital radius. Eventually its luminosity will reduce to zero and it will become an inactive black dwarf [3], [13]. In order to retain habitability, solar inso-lation on the Earth's surface must be kept at its current rate as long as possible.

**TABLE 1: Essential metallicity inventory of the solar system** [6] [7] [8] [9] [10] [11] [12]

| Body | Orbital radius (au) | Total mass (kg) | Grav. escape energy (MJ/kg) | Resource |
|---|---|---|---|---|
| Sun | 0 | $2 \times 10^{30}$ | 20,000 | H, He, metals |
| Solar wind | - | $3 \times 10^{16}$/yr | 0 | H, He, metals |
| Mercury | 0.4 | $3 \times 10^{23}$ | 9 | metals |
| Venus | 0.7 | $5 \times 10^{24}$ | 54 | metals |
| Earth | 1 | $6 \times 10^{24}$ | 63 | metals |
| Moon | 1 | $7 \times 10^{22}$ | 3 | metals |
| Mars | 1.5 | $6 \times 10^{23}$ | 13 | metals |
| Asteroid belt | ~2.7 | $2 \times 10^{21}$ | ~0 | metals |
| Jupiter | 5.2 | $1 \times 10^{27}$ | 1800 | H, He, metals |
| Jupiter's moons | 5.2 | $4 \times 10^{23}$ | ~3 | metals |
| Saturn | 9.6 | $6 \times 10^{26}$ | 650 | H, He, metals |
| Saturn's moons | 9.6 | $1 \times 10^{23}$ | ~3 | metals |
| Uranus | 19 | $9 \times 10^{25}$ | 230 | H, He, metals |
| Uranus' moons | 19 | $9 \times 10^{21}$ | ~0.3 | Metals |
| Neptune | 30 | $1 \times 10^{26}$ | 280 | H, He, metals |
| Neptune's moons | 30 | $2 \times 10^{22}$ | ~3 | metals |
| Kuiper belt | ~44 | $1 \times 10^{23}$ | ~0 | metals |
| Oort cloud | >2,000 | $2 \times 10^{25}$ | ~0 | metals |

## 2.1 STAR LIFTING

In one proposed technique to influence the Sun's evolution, a sufficiently advanced civilisation could remove mass from the Sun's surface, reducing the core pressure and slowing its rate of fusion. This could be done by using a Dyson swarm, an orbiting fleet of space-based mirrors or collectors, which captures a portion of the Sun's power output and directs it onto a small region of the solar surface. Photospheric particles would gain kinetic energy until they move faster than the Sun's gravitational escape velocity [5].

Dyson swarms may be infeasible to maintain at sufficient scale for long timeframes, due to the active guidance required to prevent a collisional cascade, as well as asteroid impacts [14], [15]. An alternative technique may be to lift the mass using space elevator technology, but this would require as-yet uninvented materials [16]. However, due to the speed of technological progress relative to the Sun's evolutionary timescale, this mass lifting technique will be nonetheless explored for completeness.

Such a process is known as star lifting (SL), first popularised by David Criswell [17]. Planet Mercury could be mined for material, which is then refined, used to manufacture solar collectors, and launched into solar orbit. As this operation progresses, the mirrors capture solar energy which is used to manufacture and launch further collectors, allowing a feedback loop of energy capture. Accordingly, such a Dyson swarm could be built in a matter of decades. This paper will take the Dyson swarm's orbital radius to be 10 $R_\odot$.

An SL rate of $5 \times 10^{19}$ kg (or $0.05M_{Ceres}$) per year was given as the optimum SL rate for life-extension by Scoggins and Kipping. Such a rate would require harnessing around 1% of the Sun's power output ($0.01L_\odot$), which would require $7 \times 10^{-3}$ of Mercury's total mass, assuming conversion of its metals into a foil with thickness 0.5 mm [18], [5].

This paper used the Modules for Experiments in Stellar Astrophysics (MESA) code to simulate the effect of SL on the future evolution of the Sun, with an SL rate of $5 \times 10^{19}$ kg per year added to the *mass_loss* variable from the present day until the core hydrogen fraction in the solar core reached 0.01 [19]. Table 2 shows the MESA simulation of the Sun's and Earth's major transition events. The SL rate of $5 \times 10^{19}$ kg per year (hereafter referred to as $\delta_{SL}$) removes a total mass of $0.13M_\odot$ over 5.1 Gyr, or $1.3 \times 10^{29}$ kg.

While SL grants a modest delay of the end of the Sun's main sequence from 4.7 to 5.1 Gyr after present, an interesting effect is that it dramatically extends the time that the Earth remains in the Sun's habitable zone (HZ) from 1.7 Gyr to 9.2 Gyr after present - well beyond the end of the Sun's main sequence (see also Fig. 1).

## 2.2 Luminosity replacement

An alternative technique which would forgo the requirement for star lifting would be luminosity replacement. To keep the insolation on the Earth within the current habitable range, the Sun's increasing luminosity can be altogether blocked from the Earth, and a replacement luminosity maintained in its place. Such a strategy would require two systems outlined here.

### 2.2.1 Sunshade

Aiming to occlude the Sun completely, the natural place to position a sunshade is the first Lagrange point L1, which is currently 0.01 au from the Earth. Combining the relation for the Earth-L1 distance with the relation for the Sun's angular width

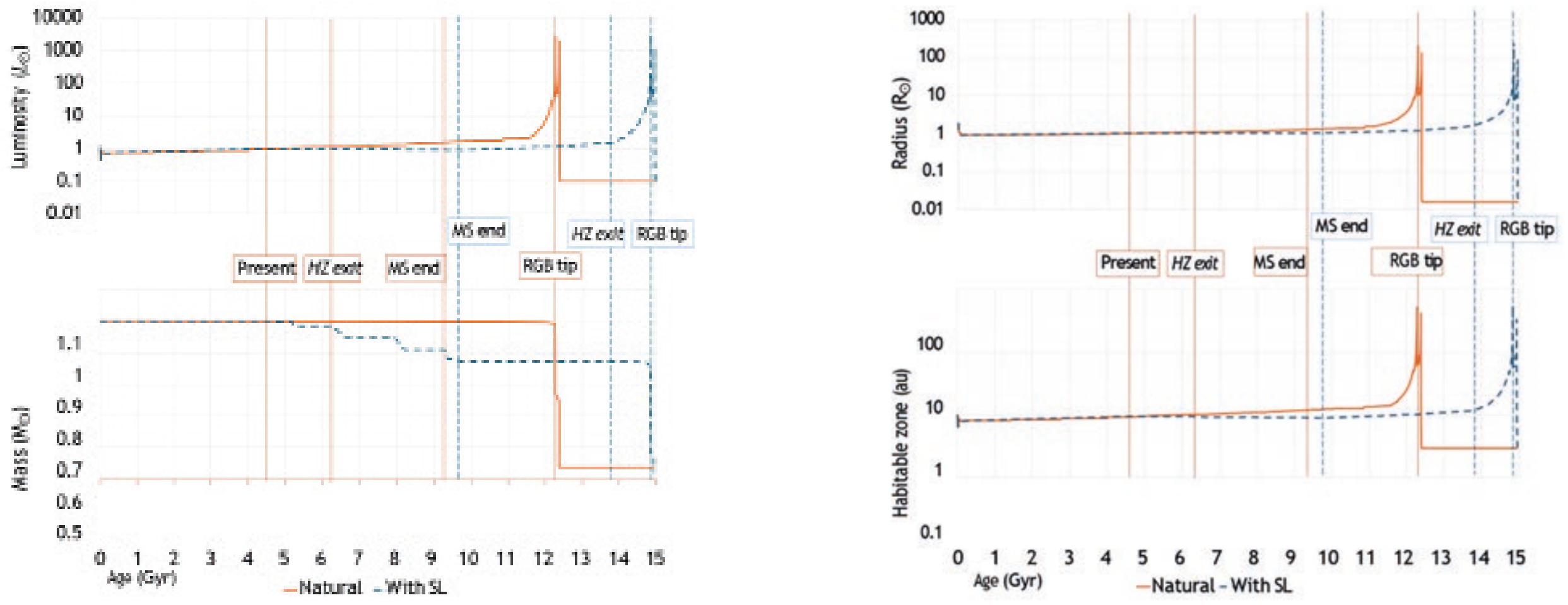

**Fig.1 Evolution of the Sun's luminosity, radius, mass and habitable zone without and with starlifting of $5 \times 10^{19}$ kg per year from the present day until hydrogen core fraction of 0.01.**

in the sky, we have for the required sunshade radius $r_{ss}$:

$$r_{ss} \approx R_{Sun} \sqrt[3]{\frac{M_{\oplus}}{3M_{Sun}}} \tag{1}$$

Per Table 2, assuming an $r_{\oplus}$ of 1.2 au before the MS end (see Sec. 3), $r_{ss}$ would be on the order of 1.5 million km. The impracticality of such a large shade is compounded by the McInnes bound: for the structure to remain in position, the gravitational force acting on it between L1 and the Sun must balance with the solar radiation pressure, and thus sets a minimum surface density which is orders of magnitude above even that of ultra-light materials such as graphene, thus eliminating any possible mass savings using these materials [21].

A technique proposed by Szapudi is to discard the restriction of the structure's centre of mass and sunshade occupying the same position – if a counterweight was positioned just beyond L1, and attached to a long tether to hold a sunshade between L1 and the Earth, this would avoid the McInnes bound restriction and achieve two orders of magnitude mass savings as compared to a structure entirely contained within L1. The technique proposed by Szapudi is for a sunshade to slightly dim the current Sun as a mitigation against climate change [21]. Scaling up this technique could provide a method to occlude the Sun during its red giant phase (RGB tip).

To this end, the distance from Earth to L1 would be 2.4 million km and 3 million km, and the angular width in the sky of the sunshade would require a minimum of 54° and 60°, for natural and SL-influenced solar evolution respectively.

By only deploying the sunshade just outside the Moon's orbit at 0.5 million km, it would be able to occlude a 70° area of the sky with a radius of 350,000 km. The gravitational pull on the sunshade from the Earth and the Moon would be negligible. The significant forces acting on the structure would be the Sun's gravity acting on the counterweight just beyond L1, and solar radiation pressure and the solar wind acting in the opposite direction on the sunshade.

**TABLE 2: MESA's simulated evolutionary transitions of the Sun and Earth***

| Transition | Age (Gyr) | Lum. ($L_{\odot}$) | Temp. (K) | Mass ($M_{\odot}$) | Rad. ($R_{\odot}$) | $r_{\oplus}$ (au) |
|---|---|---|---|---|---|---|
| **MS start** | 0 | 1.13 | 4,382 | 1.00 | 1.85 | 1 |
| **Present** | 4.6 | 1.00 | 5,760 | 1.00 | 1.00 | 1 |
| **MS end** | 9.3<br>9.7 | 1.55<br>0.88 | 5,766<br>5,519 | 1.00<br>0.87 | 1.03<br>1.25 | 1<br>1.15 |
| **RGB tip** | 12.3<br>14.8 | 2,778<br>1,429 | 3,011<br>2,374 | 0.77<br>0.59 | 194<br>224 | 1.30<br>1.69 |
| **He burn end** | 12.4<br>14.9 | 89<br>77 | 4,392<br>4,742 | 0.75<br>0.57 | 16<br>13 | 1.33<br>1.75 |
| **ARGB end** | 12.4<br>14.9 | 1,732<br>1,111 | 3,667<br>3,875 | 0.54<br>0.52 | 103<br>74 | 1.85<br>1.92 |
| **WD start** | 12.4<br>15.0 | 0.1<br>0.1 | 25,740<br>25,583 | 0.54<br>0.52 | 0.02<br>0.02 | 1.85<br>1.92 |
| **Earth HZ exit** | 6.3<br>13.8 | 1.1<br>1.5 | 5,787<br>5,040 | 1.00<br>0.87 | 1.06<br>1.61 | 1.00<br>1.15 |

*Without and with starlifting of $5 \times 10^{19}$ kg per year from the present day until hydrogen core fraction of 0.01. MS = main sequence, RGB = red giant branch, He = helium, ARGB = asymptotic red giant branch, WD = white dwarf, HZ exit = the time at which the Earth leaves the habitable zone. The current habitable zone is from 0.95 to 1.37 au [20].

Although the red giant Sun's solar wind terminal velocities will be slower than the present-day Sun (less than 100 km/s vs today's 800 km/s), the overall ram pressure will be up to $10^9$ times higher than today. Incidentally, the Earth's intrinsic magnetosphere would be too weak to protect the Earth's atmosphere from erosion from this increased ram pressure [22]. Ram pressure scales as $nV^2$, where $n$ is the particle density and $V$ is the particle velocity, implying a ram pressure of around 1-6 $\times 10^{-2}$ N/m$^{-2}$. The ram pressure on the large sunshade would be around $10^{17}$ N [23], [24]. Additionally there would be solar radiation pressure, which scales as $L_\odot/r^2$. At RGB tip, this would be $5.1 \times 10^{-3}$ Nm$^{-2}$, or $5 \times 10^{15}$ N on the sunshade.

The ram pressure can be significantly reduced or eliminated by an electromagnet placed on the Sun-facing side of the counterweight. Assuming a deflection angle of 7° and a magnetic field acting over a 1,000 km distance and accounting for bulk momentum flux, a 12.7 µT electromagnet would be able to deflect the solar wind around the solar shade.

The unavoidable solar radiation pressure on the sunshade, as well as the ram pressure if the electromagnet failed, would place an enormous force on the tether and so would require it to be made of a very strong material. Much progress has been made in the last century in maximum theoretical tensile strength, from tungsten carbide in the year 1918 (0.3 GPa) to carbon nanotubes in 2000 (63 GPa) to carbyne in 2015 (250 GPa at 77 K) [25], [26], [27]. Pessimistically assuming no further theoretical progress, the latter material would be strong enough for the tether, if manufactured at sufficient scale.

Assuming carbyne's tensile strength of 250 GPa and density of 3.5 grams per cubic centimetre, the total required mass of the tether, with a 300% safety tolerance, would be $6 \times 10^{18}$ kg. The source of this total carbon could be around 3% of the carbon-rich asteroid belt, or 40% of its dwarf planet Ceres. Alternatively, the solar wind is composed approximately 0.02% of carbon ions, thus $1.5 \times 10^{-7}$ $M_{SL}$, or 400 years of $\delta_{SL}$, would provide the necessary carbon [28] [29]. Conservatively assuming the sunshade had a total thickness of 0.1 mm (equal to the sunshield currently deployed on the James Webb Space Telescope) and a density of alumina fabric, it would have a mass of $10^{17}$ kg [30] [31]. The required fabric could be mined from $10^{-4}$ of the Moon's mass [32].

Construction of the tether could be done by unmanned drones which roam Ceres and mine, process and manufacture the carbyne. Starting with a manufacturing capacity of 1 tonne per year and initially ramping up capacity by building self-replicating construction drones, assuming a doubling time of one year to build each drone, 50 years would be enough to manufacture the required structure.

Since the L1 point is stable in the perpendicular plane but weakly unstable along the Sun–Earth axis, active station keeping is needed. As detailed by Szapudi, this can be done by altering the deflection angle of the sunshade or the sun-facing side of the counterweight [21].

The threat to the structure would be near-Earth objects and other asteroids. It would be straightforward for a civilisation capable of building the structure over long timescales to clear this debris from Earth's orbit (see Sec. 7). Nevertheless, such a safeguard would have the highest risk profile of all, as failure of the sunshade would lead to extermination of the Earth's biosphere in a matter of hours. In the event of an imminent structural, station keeping, or other failure of the sunshade, a temporary sunshade could be manoeuvred into position while a pre-made replacement was deployed at L1.

Positioned 2,000 km above the Earth, a 1 million-tonne alumina-coated graphene sunshade with an area of 200 million km$^2$ (equivalent to the Pacific and Indian oceans) could actively hover above the Earth's surface and block any radiation for one month while the main replacement was deployed at L1. The energy source for the month of hovering could be 3,400 tonnes of the antimatter production outlined in Sec. 4. Both the temporary and replacement sunshades, in folded configurations, could be waiting on the Moon to allow for low required launch energy. If the temporary sunshade was housed on the far side of the Moon, it would protect the Earth's biosphere in the case of catastrophic antimatter containment failure, which would safely be five and six orders of magnitude smaller than the Moon's orbital and gravitational binding energy respectively.

While appearing risky at first glance, any civilisation pursuing this strategy would have aeons in which to refine and test it during the Sun's main sequence. As long as near-Earth objects were adequately cleared and enough cable redundancy constructed, there would be no natural risks to the structure.

### 2.2.2 Artificial sun

With the Sun's brightening luminosity blocked from the Earth, a comfortable replacement luminosity would be required. The solar irradiance on Earth is currently 173,000 TW, or $5.4 \times 10^{24}$ J per year [33]. A possible non-solar location to source this energy would be the solar system's gas giants. Proposals have already been explored to mine these planets for hydrogen, helium-3 and helium-4 fuel [34].

Energy could be produced in situ by fusion reactors in the giant's atmosphere. To consider a fusion reaction path which would effectively produce the desired energy, it could start with the triple alpha process:

$${}^4_2He + {}^4_2He \rightarrow {}^8_4Be$$

$${}^8_4Be + {}^4_2He \rightarrow {}^{12}_6C + 2\gamma$$

which releases approximately 7.3 MeV.

As suggested by Caplan [34], hydrogen can be fused onto the produced $^{12}C$, as in the first stages of the hot CNO cycle,

$${}^{12}_6C + {}^1_1H \rightarrow {}^{13}_7N + \gamma$$

$${}^{13}_7N + {}^1_1H \rightarrow {}^{14}_8O + \gamma$$

which releases 1.9 MeV and 4.6 MeV respectively.

The produced oxygen then beta decays into nitrogen,

$${}^{14}_8O \rightarrow {}^{14}_7N + e+ + \nu e.$$

The resultant positrons can then annihilate with the electrons produced during the ionisation of the plasma before the start of the process (the energy of which will be ignored for simplicity).

The net process would be:

$3\,{}^4_2He$ and $2\,{}^1_1H$ producing $1\,{}^{13}_7N$, or 3 × 4.0026032 u

+ 2 × 1.0078250 u = 14.0234596 u input vs. 14.003074 u output, a mass defect of 0.0203856 u. The energy density would therefore be $1.3 \times 10^{14}$ J $kg^{-1}$, with roughly a 3:1 fuel mix ratio of helium-4 to hydrogen. The fusion reactor would require a temperature on the order of $10^8$ K [36]. By diving to 6,500 km depth on Jupiter, the reactors would experience 4 GPa of inward ambient pressure, balancing the huge out - ward pressure of the required magnetic confinement during fusion. MHD confinement could be done by fast-circulating liquid lithium.

The solar system's gas giants collectively contain $3.8 \times 10^{26}$ kg of the aforementioned fusion fuel, or $4.9 \times 10^{40}$ J of available fusion energy [37] [38], [39], [40], [41]. If this energy was produced and beamed onto the Earth's surface to replicate the current natural solar wavelengths, it could provide $9.1 \times 10^{15}$ (9.1 million billion) years of sunlight to the Earth.

Theoretically this could be increased by gradually depositing interloping interstellar medium onto Jupiter. The Birch ring can run a circular current which produces an artificial magnetosphere around Jupiter. An array of lasers firing the resonant 243 and 58 nm wavelengths for hydrogen and helium would be able to ionise the interstellar medium flowing near Jupiter, causing it to spiral down onto Jupiter via the Lorentz force. Ionising a 5.7 million km radius volume, this mechanism would capture 300 kg of fuel per second and consume 3% of the 173, 000 TW produced by fusing the same amount. The spent fuel could be lifted off Jupiter via the Birch ring with a negligible energy expenditure. This mechanism would keep replenishing fuel while there is still interstellar medium available. After the Milky Way - Andromeda merger, 70 billion solar masses of medium will be available to be harvested during Earth's orbit around the galactic centre, extending the potential habitability time to 14.8 nonillion years.

Floating in Jupiter's atmosphere, the fusion reactors would collectively breathe in 300 kg of helium 4-hydrogen mix per second and breathe out roughly the same mass of nitrogen.

In the fusion aerostats, the produced energy could then be beamed via microwaves to a generator at the giant's L1 point, in order to avoid the inefficiencies of atmospheric energy absorption. At the L1 generator stage, the energy is converted to the natural spectrum of sunlight (without the biologically harmful UV-B and -C rays) in the form of a large laser array, and then beamed to the Earth. To deliver the necessary power, a laser array with a diameter of 15 km would require a power density of around 1 GW $m^{-2}$, which is achievable for modern lasers [44]. Per Table 1, each gas giant's moons provide an ideal staging location for construction and maintenance of the system, providing nearby large masses of metals with low gravitational escape energy.

In order to retain constant line of sight onto the Earth, and to provide redundancies in case of failure, the giant and the Earth would have light relays at their respective L4 and L5 points. If the L1 generator fails, the aerostats could switch to producing natural spectrum light and beam directly to an L4 or L5 relay. If Earth's L4 fails, the light could be beamed to its L5 relay instead. The relays could reflect the light beam using photonic crystal technology, which demonstrates high reflectivity but with lower material fatigue compared with traditional mirrors [45]. Assuming the relay at L4 operates continuously, the Earth's day/night and seasonal cycles would be retained as normal (however, an extra step would be required to account for Earth's eventual obliquity changes – see Sec. 6).

## 3 ENGULFMENT

### 3.1 Star lifting

At RGB tip, the Sun's radius will balloon to its peak radius and threaten to engulf the Earth. While the Earth's orbital radius $r_\oplus$ would increase due to the Sun's mass loss, it likely would not be enough to escape engulfment. In the literature there are varying figures for the solar radius at RGB tip. A 2009 simulation computed an RGB tip radius of 146 $R_\odot$, while the Modules for Experiments in Stellar Astrophysics (MESA) code computes 194 $R_\odot$ [46]. A 2008 paper by Schröder and Smith concluded that the Earth would not quite escape ultimate engulfment due to tidal forces imparted by the red giant's tidal bulge and increased radius, which was computed as 256 $R_\odot$, with solar mass loss expanding Earth's orbital radius to 322 $R_\odot$. The torque exerted on the Earth at RGB tip, due to the slowing of the Sun's rotation and subsequent tidal bulge formed by the Earth's gravitational pull, is expressed as:

$$\Gamma = 6\frac{\lambda_2}{t_f} q^2 M_{Sun} R_{Sun}^2 \left(\frac{R_{Sun}}{r_E}\right)^6 (\Omega - \omega) \tag{3}$$

where $\lambda_2$ is a convective envelope coefficient, $t_f$ is the convective friction time, $q$ is the mass ratio $M_{Earth}/M_{Sun}$ at time $t$, $M_{Sun}$ and $R_{Sun}$ are the mass and radius of the Sun at RGB tip, $r_E$ is the orbital radius of the Earth at RGB tip, $\Omega$ is the angular velocity of the solar rotation, and $\omega$ is the angular velocity of the Earth [3]. This yielded an angular decay time of the same order of magnitude as the time the Sun will spend near RGB tip. The Earth's loss of angular momentum, further accelerated by subsequent dynamical friction in the lower chromosphere, would result in engulfment.

SL has a net effect on the RGB tip radius. With SL removing mass from the solar envelope rather than the core, there is an interplay between the reduced gravitational pressure vs decreased pressure on the core resulting in lighter hydrogen shell burning. Ultimately SL results in a 15% gain in RGB tip radius contrasting with a 23% loss in RGB tip mass and resultant increase in $r_\oplus$. From Equation 1 we can estimate the order of magnitude effect that this would have on the tidal bulge's torque on the Earth as:

$$\Gamma \propto -{L_{Sun}}^{1/3} M_\oplus^5 M_{Sun}^{6\frac{2}{3}} R_{Sun}^{7\frac{1}{3}} \tag{4}$$

where $M_\oplus$ is the mass of Earth. This gives an estimate of the Earth in the natural MESA simulation experiencing 66% less typical torque than in the 2008 analysis, and in the SL simulation experiencing 89% less typical torque. The Schröder and Smith paper gives an orbital momentum decay time of $2.6 \times 10^6$ years, which is comparable to the time that the Sun spends near RGB tip – the Sun being larger than 100$R_\oplus$ for $3 \times 10^6$ years. It can be estimated that SL would allow the Earth to narrowly prevent engulfment – however, clearly any civilisation on Earth would want to ensure it escapes engulfment with a clear margin of safety. Therefore, artificially increasing Earth's orbital radius long in advance would be unavoidable.

## 3.2 Orbital engineering

Schröder and Smith concluded that a present Earth orbit of 1.15 AU would allow the Earth to survive engulfment due to the Sun's tidal effects [3]. The Earth therefore would require orbital energy to be transferred to it in order to increase its orbital radius before the end of the Sun's main sequence.

Korykansky et al proposed a method of using a Kuiper Belt object of mass $10^{19}$ kg, approaching at 40 km $s^{-1}$, and have it pass 3,600 km from the Earth's surface, transferring orbital energy to the Earth. Accurate timing would allow the object to cross the orbit of Jupiter on its outbound trajectory, regaining the energy it loses. $10^6$ encounters (one Earth flyby every 6,000 years) would mean a cumulative flyby mass of 1.5 $M_\oplus$, increasing the Earth's orbital radius to 1.5 au over 5 Gyr, ensuring constant solar energy incident on Earth as the Sun's luminosity increased [47]. The energy expenditure of the scheme, corresponding to the kinetic energy of the object at 50 km $s^{-1}$, is an inexpensive $10^{29}$ J.

A potential drawback of the scheme given by Korycansky *et al.* is that such an object passing so close to Earth would impart around 10 times the Moon's tidal force. Such a force would lead to a spin up of the Earth's rotation, although this could be cancelled out by proper timing of the outgoing encounters. Tidal effects may also lead to negative localised effects such as coastal flooding. The flyby may also destabilise the Moon's orbit and so would require a supplementary flyby to correct it. Increasing the Earth's orbit to 1.5 au would also interact with Mars' orbit and traverse secular and mean-motion resonances with other planets. The overriding major catastrophic risk as emphasised by Korycansky et al. is an error in trajectory of the object, leading to it impacting the Earth's surface which would destroy the biosphere, at least down to the level of bacteria.

A more energy-intensive but potentially safer modification of this scheme would be to fire a particle beam close to Earth to provide the flyby mass. Instead of a large flyby mass every 6,000 years, the beam could provide a stream of much smaller mass continuously. This scheme would avoid the tidal disruption and all-or-nothing risk of a large object flyby. Corrections to the Moon's orbit and the Earth's rotation could also be readily done by re-aiming the beam.

This technique could be used for engulfment avoidance only – thus for Earth's orbit to increase to 1.2 au, to exceed the 1.15 au minimum safe distance as calculated by Schröder and Smith [3]. This would avoid the problems of potentially disturbing Mars' orbit. This new orbit of Earth was simulated with the REBOUND N-body integrator code and was found to produce no instabilities in the solar system during a full 100 Myr simulation [42].

Extending Earth's orbit to 1.2 au would require $3.5 \times 10^{32}$ J. Per Korycansky et al.'s formulations, in the heliocentric frame, the flyby mass has a tangential velocity matching Earth's orbital velocity. The energy transfer $\delta Q$ per unit mass $m$ is:

$$\frac{\delta Q}{m} = V_\oplus (v \sin \vartheta) \qquad (5)$$

where $V_\oplus$ is the Earth's orbital velocity, $v$ is the incoming relative speed (purely radial) and $\theta$ is the deflection angle. For a hyperbolic flyby:

$$\cot\left(\frac{\theta}{2}\right) = \frac{v^2 b}{\mu_\oplus} \qquad (6)$$

where $\mu_\oplus$ is the gravitational constant multiplied by Earth's mass, and $b$ is the encounter's distance from the Earth's centre which we will set to 10,000 km. The particle beam will simply dissipate after the flyby. Thus the maximum energy transfer occurs when x = 1 (i.e., θ = 90°, where sin θ = 1):

$$\frac{v^2 b}{\mu_\oplus} = 1 \implies v_{optimum} = \sqrt{\left(\frac{\mu_\oplus}{b}\right)} \qquad (7)$$

giving an optimum approach speed of 6.3 km $s^{-1}$, which per Equation (5) gives an energy transfer efficiency of $1.880 \times 10^8$ J $kg^{-1}$ implying a total required flyby of 0.31 Earth masses.

The choice of mass could be Jupiter's oxygen reserve due to the element's environmental safety if accidentally beamed at the Earth – Earth's atmosphere holds $10^{18}$ kg of oxygen already and oxygen is a non-toxic, non-greenhouse gas. Jupiter is estimated to hold 4.6 to 9.2 Earth masses of oxygen in its deeper regions [48], which could be accessed by the aerostats described in Sec. 2.2.2 by altering their buoyancy.

The oxygen-rich depth is around 100 km, which would have an equivalent pressure of 70 m under Earth' s oceans. Due to zonal jet speeds at this depth, the mass flux could be captured by a million-strong fleet of stationary 30 by 30 m scoops. Once gathered, the mass is then lifted to a Birch orbital ring circling the planet, which could be made from current materials and sourced from a small satellite asteroid such as Metis [90]. Running along the Birch orbital ring's tracks, the oxygen mass is then launched in a Hohmann transfer to a catcher-launcher station at Earth's L2 point.

The mass would arrive tangentially to Earth's L2 point at 8.7 km $s^{-1}$, where it would be caught and then launched at 4.7 km $s^{-1}$ so it crosses Earth's leading edge at the optimal 6.3 km $s^{-1}$. A small mass of helium could also be launched to power the launching and any station keeping.

Taking place over 100 Myr, assuming the Earth will be out of the beam's range for half of Earth's year, 120,000 tonnes of oxygen per second would be flying by, equivalent to around half of the Amazon River's outflow. The beam would be safe if accidentally aimed at the Earth due to its spread of 1,000 km, which would only hit the atmosphere with the same areal kinetic energy per second as a thunderstorm [49].

To compensate for the reduced insolation on the Earth due to its expanding orbit, as in the luminosity replacement system in Sec. 2.2.2, fusion aerostats in Jupiter's atmosphere could produce and beam energy to a relay at the Earth's L1 point, which could power a laser to shine natural-equivalent sunlight onto the Earth to provide any supplemental insolation (see Fig. 2).

The total energy required to lift the oxygen mass from Jupiter would be equivalent to $10^{19}$ kg of the helium fuel described in Sec. 2.2.2, or $10^{-7}$ of Jupiter's helium fuel reserve.

# 4 END OF TECTONICS

Earth's current internal heat energy is 47 TW, equivalent to 0.03% of the solar insolation on the surface [33]. It is the driv-ing force of tectonics, and is caused by radioactive decay in the mantle and crust, as well as primordial heat left over from the Earth's formation [50], [51]. Tectonic activity is important for the Earth's biosphere, as it recycles vital elements, regulates the atmosphere, and causes secular changes to the surface (mountain building and so on), creating varieties of habitats which drive biodiversity [52], [53], [54].

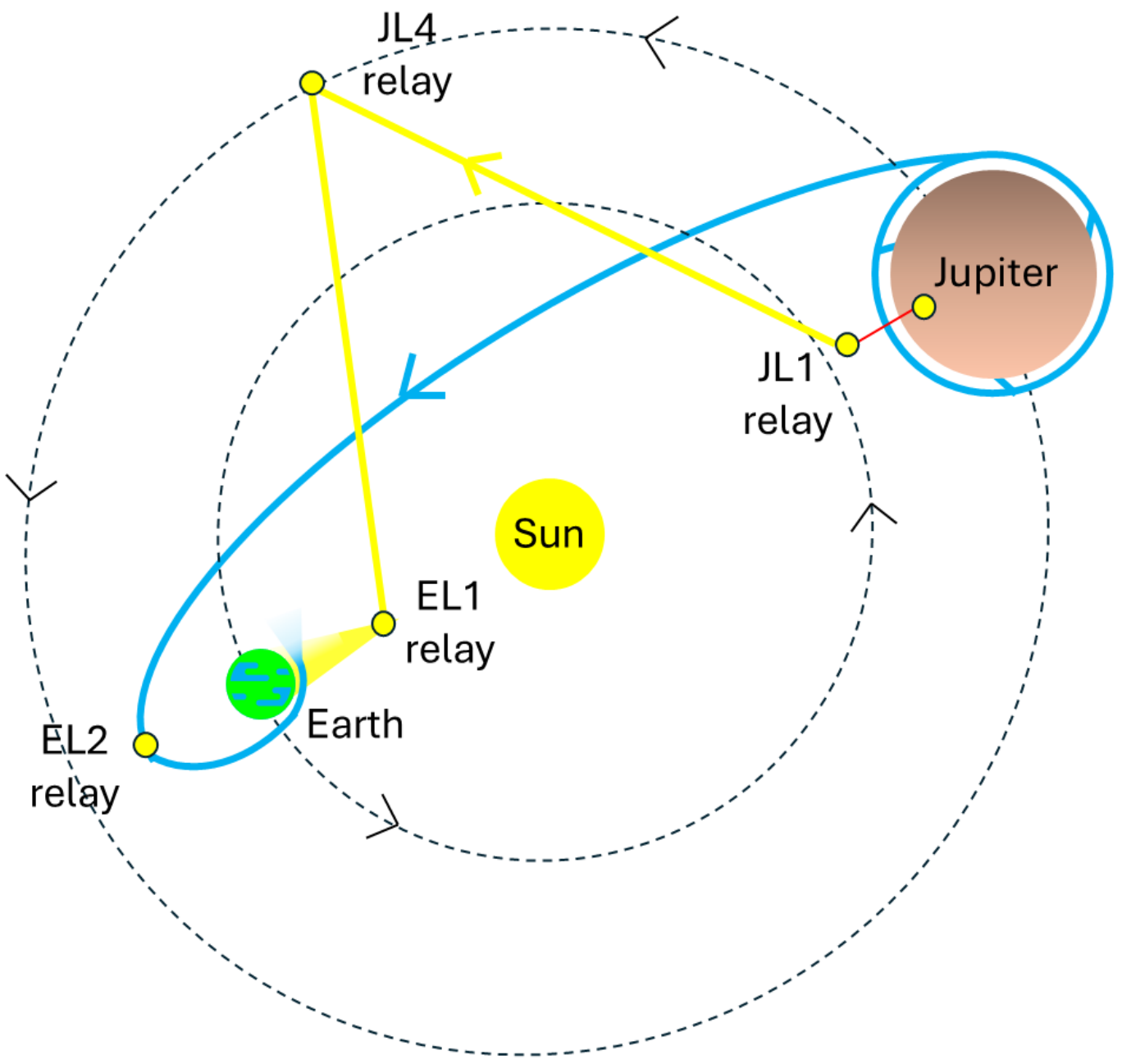

**Fig.2 Essential diagram of an orbital engineering system. Oxygen is gathered from Jupiter's deeper regions by variable-buoyancy aerostats, then lifted to a Birch orbital ring circling the planet. At the appropriate launch window, it's then launched in a Hohmann transfer to a catcher- launcher station at Earth's L2 point. It is then launched at the Earth's leading edge, with the gravitational deflection of the beam imparting energy to the Earth's orbit and gradually increasing its orbital radius, to avoid eventual engulfment by the red giant Sun in around 8 Gyr. The beam's 1,000 km spread, as well as its atmospherically benign oxygen content, would mean negligible safety risks if mis-aimed at the Earth. To compensate for reduced insolation due to the Earth's expanded orbit, supplemental energy could be produced by helium 4 and hydrogen fusion in Jupiter's atmosphere, beamed to a relay at the Earth's L1 point, and then shone onto the Earth along with the natural sunlight from the main sequence Sun. This infrastructure could also readily provide water to Mars for terraforming purposes.**

However, due to the gradual cooling of the Earth's interior, plate tectonics will end in around 1.45 Gyr [55]. In order to retain this process, it would be necessary to replicate the current naturally-produced 47 TW with an artificially-sourced heat flux. For the heat source to output isotropically from the interior, mimicking the natural process, it must be placed as deep as possible, either in the core or at the core-mantle boundary.

As a method to investigate the Earth's interior, Stevenson proposed blasting open a crack in the Earth's crust and pouring molten iron into it, in order to send a probe to the Earth's core-mantle boundary, with the probe communicating readings back to the surface via acoustic vibration. A multiple megaton nuclear device ($10^{15}$ J) would be detonated to open a 0.1 by 300 m crack, which would be the equivalent of a magnitude 7 earthquake (of which there are naturally around 15 every year) [56]. Into it would be poured $10^4$ m$^3$ ($10^8$ kg) of molten iron, equivalent to around an hour's production of the world's foundries. As the iron mass sinks, the crack would close behind it, and it would reach the Earth's core in around a week [57]. Within the molten iron would be the probe, encased in a high melting point alloy, in saturation equilibrium with the iron.

In order to maintain Earth's tectonics, this scheme could be modified by replacing the molten iron with a radioactive isotope with a suitably high heat output, half-life, and production efficiency, with the isotope's heat output providing the required energy. However, even with retrieval of upwelled material, over long timescales this would be unsustainable due to some material remaining inaccessible and accumulating deep in the interior. Instead, due to its favourable energy density and energy discharge purity, the energy could be provided via annihilation of antimatter.

An injected antimatter mass of 8.2 tonnes per year would be required to replicate the 47 TW. In order to minimise the risk of catastrophic containment failure, small amounts of antimatter could be produced, contained and injected through the crust regularly, by production facilities situated at remote, geologically favourable locations which could forgo the need for large detonations to initiate the crack each time [57].

Frozen pellets of antihydrogen would be the most feasible to produce and store safely. Cooled to 0.5 K, such pellets could be magnetically confined in a vacuum container using their dipoles [58]. Once held inside the molten iron and injected into the mantle, the container would purposely have a week of onboard energy to confine the antimatter. After the container reaches the core-mantle boundary, this energy would be exhausted, the containment system would cease to work, and the antimatter would contact the interior of the container, converting the mass into gamma rays, which would be absorbed as heat by the iron and surrounding rock.

Running continuously, ten facilities dotted around the globe would each need to produce and sink 2.2 kg of antimatter daily. Although currently expensive to produce, mass production of antimatter has not seriously been attempted, with production confined to byproducts from particle experiments. Nevertheless, since 2014 worldwide production has been progressing at a virtually geometric rate. It is forseeable that given enough time and attention, the required production rate could be possible in the future [58].

Sudden energy releases by the annihilation process hold the risk of extreme thermal expansion of the molten iron and the surrounding rock, which could produce dangerous earthquakes on the surface. Fortunately, the blob of molten iron which surrounds the container would be a good shock absorber for this purpose. Due to the high pressure at the core-mantle boundary, the iron would not vaporise unless heated to on the order of $10^4$ K, and iron's high specific heat capacity and low thermal expansion coefficient mean the blob of iron would heat up without a shockwave-producing expansion. Embedding ten small antimatter containers in the molten iron and staggering their annihilation to every 2.4 hours, each event would have a moment magnitude of 4 which would be imperceptible on the surface [59].

This process could continue essentially indefinitely. The mantle replenishes iron into the crust at the rate of 21 $km^3$ per year at mid-ocean ridges alone, equating to $5 \times 10^{12}$ kg of iron replenishment each year [60]. Assuming 10% efficiency, the antimatter production would require capturing 0.6% of the solar insolation on the Earth's surface, or a cumulative global solar capture of $3 \times 10^6$ $km^2$, equivalent to the area of India. Over the nonillions of years of sunlight production detailed in Sec. 2.2.2, the container material could be supplied via capturing metals from the interstellar medium, using the same mechanism as for the hydrogen and helium.

## 5 WATER LOSS

The Earth's upper atmosphere is currently losing mass due to three mechanisms: charge exchange escape, where ultraviolet sunlight produces fast ions which collide with neutral atoms and capture an electron, resulting in fast neutral atoms (60-90%); thermal (Jeans) escape (10-40%), and the polar wind, where the solar wind drags magnetic field lines above the poles and into space, allowing ions to escape (10-15%). In total this results in $10^{25}$ oxygen atoms lost per second, or 0.3 kg $s^{-1}$. of water loss from the Earth [61].

Currently the Earth's total water reserve is $1.4 \times 10^{21}$ kg, or $2.3 \times 10^{-4}$ $M_\oplus$, with 97% of it held in its oceans [62]. Assuming implementation of the luminosity replacement system outlined in Sec. 2.2, the current ultraviolet and thermal effects would mostly continue as normal, maintaining around the current rate of water loss. In a simple linear model of future water loss, Earth would thus hold enough water for $1.4 \times 10^{15}$ years. In order to match the full lifetime of the artificial sun described in Sec. 2. 2. 2, an average 1,000 tonnes per year of water would have to be added to the Earth from elsewhere.

This water could be readily provided using a fraction of the particle beam outlined in Sec. 3.2, by mining the 1$M_\oplus$ of water in Jupiter's atmosphere [48]. This beam could also be used to introduce water to Mars for terraforming purposes.

Over the nonillions of sunlight production detailed in Sec. 2. 2.2, 28 grams of oxygen per second could be captured from the interstellar medium using the same mechanism as with the hydrogen and helium, and combined with hydrogen to maintain the Jovian water supply indefinitely.

## 6 AXIAL TILT AND ROTATION

Currently, the Earth's tidal bulge is exerting a gravitational pull on the Moon, increasing the Moon's distance by 38 mm and the Earth's day by 17 microseconds per year. Left unchecked, this would eventually lead to the mutual tidal locking of the Earth and the Moon in around 50 Gyr (after the Sun has faded to a white dwarf), effectively meaning that one Earth day would last more than a month [65], [66], [67]. The recession of the Moon may also diminish its stabilising effect on the Earth's axial tilt (obliquity), and it may reach as high as 90 degrees, although more recent modelling has indicated this may not occur [68] [69]. Assuming that insolation on the Earth's surface continued to shine from a constant ecliptic direction, such changes would be disastroud for the Earth's habitability [70].

The solar insolation on the Earth effectively rotates about it once every 24 hours, with a yearly seasonal oscillation of around 250 to 480 W per sq m for each hemisphere [71]. To replicate this cycle, the artificial sun could orbit over the equator with radius $r$. Taking Kepler's third law and interpreting the period as the difference between the angular velocities of the orbiter and the rotating Earth, we have:

$$r = \left( \frac{G\, M_\oplus}{4\, \pi^2} \left( \frac{p\, t_g}{p \pm t_g} \right)^2 \right)^{\frac{1}{3}} \quad (8)$$

where $G$ is the gravitational constant, $M_\oplus$ is the mass of the Earth, $p$ is the period of rotation of the Earth (which will increase until the mutual tidal locking in 50 Gyr), and $t_g$ is the ground track period of the artificial sun, in this case 24 hours, with the plus sign used for a prograde orbit and the minus sign for retrograde.

Occupying this equatorial orbit would give the artificial sun a view over the entire Earth's surface once every 24 hours, regardless of its rotation period or obliquity. Receiving a beam of natural-equivalent sunlight from the relay at L4, it would reflect and spread it over the required angle. Additional artificial suns, orbiting at slightly different altitudes and inclinations, could be selected depending on the time of year, in order to replicate the seasons and to provide redundancy.

Depending on the Earth's pole orientation, there may be sections of the artificial sun's orbit which would be blocked from the L4 relay by the Earth. An additional problem is safety – at a certain time of day, when the artificial sun is between the L4 point and the Earth, misaiming the beam would mean a concentrated beam of sunlight incident on the Earth's surface, which would be catastrophic. To solve this blocking and safety problem, twice per day the beam from L4 could be reflected onto the artificial sun via an additional relay placed at the L2 point – once when the artificial sun is between L4 and the Earth, and once when the artificial sun is occluded from L4 by the Earth. The switch from beaming directly from L4 and reflecting off L2 would be done gradually to account for the five second delay for the light beam to reflect off L2. Depending on the Earth's future pole orientation, there may be aeons when this step is unnecessary, when the relay at L4 has full line of sight onto the artificial sun's orbit for the full 24 hours (see Fig. 3). In order to retain the present-day Sun's angular width in the sky and its soft-edged shadows on the Earth, the artificial sun could have a diameter on the order of 10 km.

## 7 OBJECT COLLISION

There are over 37,000 near-Earth objects (NEOs), defined as those with perihelions of less than 1.3 au [72]. Such objects almost all originate from within the solar system, with around $10^{-5}$ originating from outside it [73]. Each object possesses varying degrees of threat to the biosphere of Earth, depending on various factors, primarily on the probability and kinetic energy of collision [74]. Safeguarding against NEO threats is already feasible, with an asteroid already having been successfully deflected [75], [76]. Along with safeguarding the Earth, experience in manipulating NEOs could also unlock their valuable resources in asteroid mining operations [77].

Another potential collision risk over long timescales is instabilities in Mercury's orbit, which has a small probability of a decrease in its eccentricity causing a transfer of angular momentum from the giant planets which then destabilises all the terrestrial planets, with potential planetary collisions with the Earth in 3.3 Gyr [78]. Until Mercury is engulfed by the Sun in its red giant phase, it thus presents a small but potentially catastrophic threat, while conversely presenting a large reserve of metallic resources. Mercury has relatively low gravity and no atmosphere – thus, mining infrastructure on the planet could serve the dual purpose of a mass driver, accelerating low-val-ue elements off the planet in order to provide Δv at perihelion or aphelion to correct any decreases in eccentricity. Ejecting 0.06% of the planet's mass at 100 km $s^{-1}$ would provide the necessary 100 m $s^{-1}$ of Δv to alter any changes to eccentricity.

## 8 EXTRA-SYSTEM THREATS

The solar system does not exist in isolation, but orbits the galactic centre among billions of other stellar systems and objects. Due to the immense distances, such phenomena do not pose an immediate threat, however over long timescales this threat would become significant enough to necessitate mitigation strategies. The two types of extra-system threats are explosions and encounters.

A nearby supernova, gamma ray burst or magnetar explosion, if sufficiently powerful and close, would be catastrophic for Earth's biosphere. However, as detailed by Círcović and Vukotić, these explosions would be feasible to protect against by positioning a shielding swarm of small objects or particles between the Earth and the event, held together by an electromagnetic field. Such astrophysical threats would, in principle, be predictable well in advance and thus feasible to safeguard against [79].

Finally, in the encounter threat there exists a 1 per cent chance per Gyr of a star approaching within 100 au of the Sun, which could lead to orbital instabilities in the solar system - with this probability increasing during the Milky Way-Andromeda galactic merger taking place in 4.5 Gyr [80], [81]. While a direct collision is improbable, the severity of the threat increases with the potential encounter body's mass and flyby distance, with a 10,000 au encounter triggering a comet shower in 0.18 Myr, and encounters less than 100 au becoming even more catastrophic.

While the involved distances and masses seem insurmountable, such encounters would be predictable well in advance. A safeguard against this threat would therefore be to impart an accelerative force on the Sun for a sufficiently long timespan in order to alter its galactic orbit, putting sufficient distance between the incoming encounter's closest approach and the solar system's outer reaches. The rest of the solar system would remain gravitationally bound to the Sun and move with it as normal. Various designs of these techniques, dubbed stellar engines, have been proposed by Shkadov, Fogg, Caplan, Svonoros and Vidal, which either passively reflect luminosity or actively use star lifted fuel to generate thrust [82]. Producing accelerations of the Sun from $10^{-12}$ to $10^{-6}$ m $s^{-2}$, over long enough timescales they could even be used for interstellar or intergalactic migration.

For the purposes of encounter avoidance only, a new design of stellar engine is necessary which would be able to impart an accelerative force with or without stellar luminosity or fuel, allowing the Sun, active or not, to be deflected at any time during the full order of $10^{17}$–year timescale of the set of safeguards. Using the technique described in Sec. 3.2, low-value mass from Jupiter could be gathered, accelerated, collimated and aimed just above or below the Sun's north or south pole throughout the Jovian year, using the resultant gravitational slingshot effect to gradually accelerate the Sun over long timescales along the ecliptic polar axis.

Incidentally, while the Oort Cloud offers enough mass and with negligible gravitational escape energy, this energy saving would be eliminated by the travel distance needed to capture the mass within the huge volume [63].

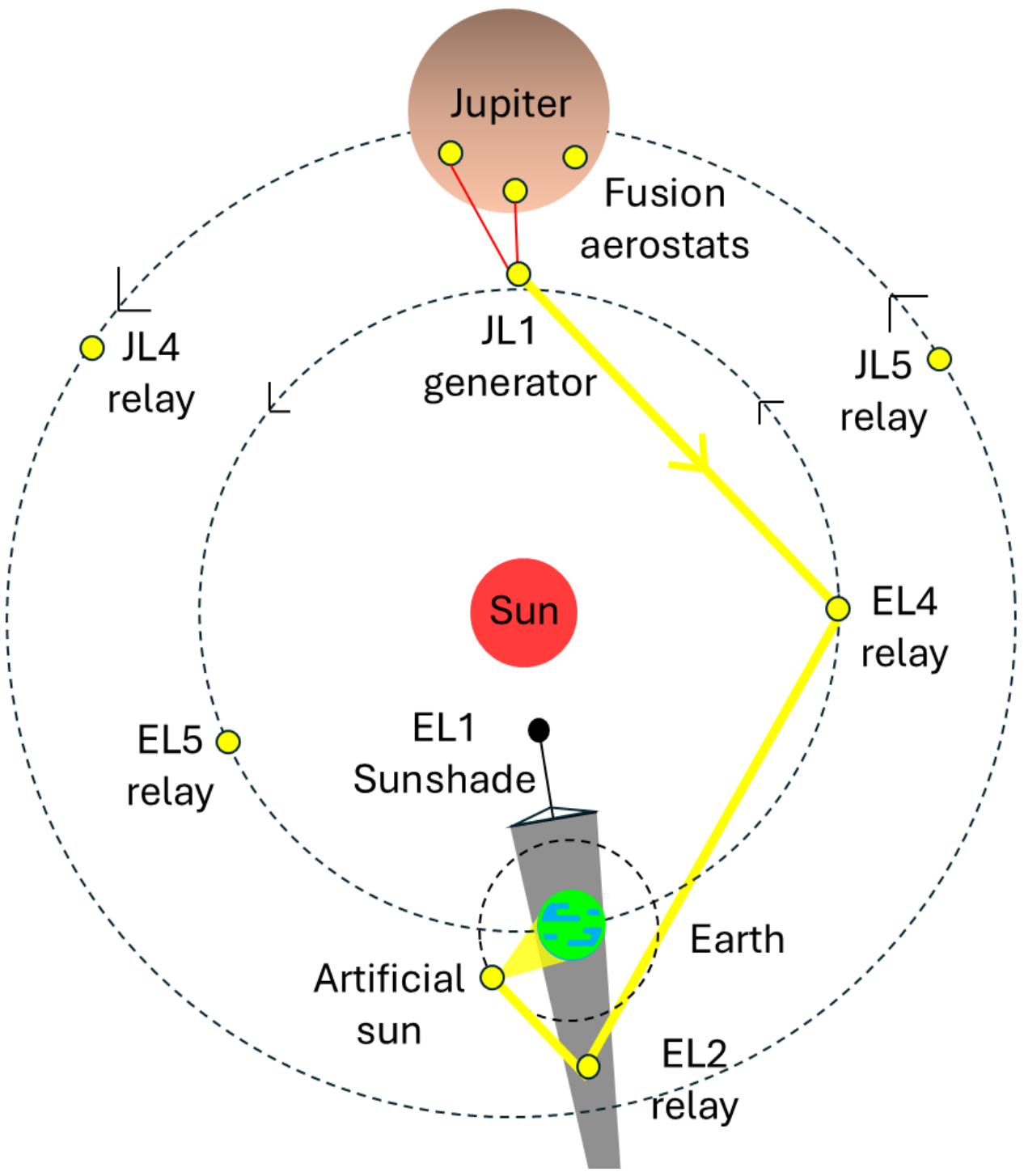

**Fig.3 Essential diagram of the luminosity replacement system. A sunshade at the Earth's L1 point occludes sunlight, while helium and hydrogen fusion in the atmosphere of a gas giant (Jupiter, Saturn, Uranus or Neptune) produces energy which is then converted to natural-equivalent sunlight and beamed onto the Earth's surface. Relays at the giant's and the Earth's Lagrange points ensure a constant flow of energy regardless of the relative positions of the giant and the Earth, or, the axial tilt and rotation period of the Earth. The system would be able to provide sunlight to the Earth's surface regardless of the Sun's evolutionary stage.**

As in Sec. 3.2, abundant hydrogen from the Jovian atmosphere could be gathered, accelerated and fired just above or below one of the Sun's poles. A total of around 5 Earth masses (2% of Jupiter's hydrogen mass) approaching at 616 km $s^{-1}$ would impart the required $10^{30}$ J of energy to deflect the solar system by 100 au and dodge the incoming encounter. Launching from Jupiter's L1 point with a radial velocity of 19 km $s^{-1}$ would ensure correct flyby speed.

Over 1 Myr, this would require $10^{12}$ kg of hydrogen per second, which would be relatively easy to access – a fleet of one million cruisers orbiting 50 km above the 1 bar surface at 40 km $s^{-1}$ would be able to capture the required mass flux using scoops 30 × 30 m in size. The total energy required to lift the mass would be equivalent to $10^{20}$ kg of the helium fuel described in Sec. 2.2.2, or $10^{-6}$ of Jupiter's helium fuel reserve.

These parameters would change depending on the trajectory of the encounter relative to the orientation of the ecliptic during the Sun's galactic orbit, which currently has a period of 225 Myr.

## 9 DISCUSSION

For any civilisation present on Earth when the Sun's luminosity begins to threaten habitability, the alternative to implementing the set of safeguards would be to leave Earth to live in space habitats, or migrate to another stellar system. The safeguard, space habitation, and migration strategies can thus be compared.

**TABLE 3: Energy estimates for implementing the set of safeguards vs humanity's energy production vs migrating $10^{10}$ humans to another solar system [83]**

| Safeguard | Resource location | Energy requirement estimate (log10) |
|---|---|---|
| Migrating civilisation | | 31 J $yr^{-1}$ for 1 yr |
| Encounter avoidance (Sec. 8) | Jupiter's hydrogen | 0 to 28 J $yr^{-1}$ for 1 Myr |
| Orbital engineering,(Sec. 3) | Jupiter's oxygen | 25 J $yr^{-1}$ for 100 Myr |
| Artificial sun (Sec. 2, Sec. 6) | Jupiter, Saturn, Uranus, Neptune | 24 J $yr^{-1}$ indefinitely |
| Projected humanity energy production in year 2400 | | 24 J $yr^{-1}$ |
| Object collision prevention (Sec. 7) | Near-Earth objects, Mercury | 0 to 22 J $yr^{-1}$ for 4 Gyr |
| Antimatter sinking (Sec. 4) | Antimatter production on Earth | 21 J $yr^{-1}$ indefinitely |
| Current humanity energy production | | 21 J $yr^{-1}$ |
| Sunshade (Sec. 2) | Ceres or asteroid belt | 18 J $yr^{-1}$ for 4 Gyr |
| Water mining (Sec. 5) | Jupiter's water | 15 J $yr^{-1}$ indefinitely |

### 9.1 Energy and economics

The energy cost of migrating an entire civilisation could be estimated as $10^{31}$ J, corresponding to the kinetic energy required to accelerate to 0.1$c$ a set of craft holding $10^{10}$ humans, using ten times the International Space Station's mass-capacity ratio ($10^{16}$ kg), and then decelerate them at the destination - although this mass ratio may be a conservative estimate for a multi-generational interstellar journey. Terraforming at the destination stellar system would need to be done millennia in advance, but would require orders of magnitude less energy [5]. As a long-term strategy, assuming the migration was to a young Sun-like star, the settlers may be able to inhabit the destination for 10 Gyr before the end of the star's main sequence, whereupon they would need to migrate again.

The alternative of remaining in the Sol system, with the implemented safeguards, would require $5.4 \times 10^{34}$ J per 10 Gyr. However, while the safeguards are more expensive in terms of total energy, the peak energy production rate would be lower, since the $10^{31}$ J produced for migration would need to be discharged as acceleration and deceleration burns. These could last as brief as 36 days each, for an acceleration of 1g. In contrast, restricting the energy discharge rate to that of the safeguards would mean burn times of a thousand or a million years (depending on if encounter avoidance is needed or not).

The set of safeguards would require mining various bodies in the solar system. The infrastructure required to do so would be able to provide an abundance of valuable resources to Earth and so would be economically viable in the near term. Implementing the set of safeguards could therefore be done by repurposing existent space mining infrastructure, the economic benefits of which are already being considered [77], [34]. In contrast, taking a short-term viewpoint, building and transporting terraforming apparatus to a distant planet before migrating to it would present no economic benefits to the home civilisation for many generations.

Future civilisations could build and inhabit large space habitats, with gravity provided by spinning the habitat. Various designs of these space stations have been proposed by Tsiolkovsky, O'Neill and others [84]. To provide energy, their orbits could be maintained inside the future Sun's evolving habitable zone, or their shells enclosed around artificial sunlight – energy could be provided by the same method as the safeguards' artificial sun.

### 9.2 Feasibility

The Earth may be exceptionally rare in its ability to host a biosphere over geologic timescales – it is in a relatively stable zone in the galaxy, away from regions of high radiation or supernovae; its host star is long-lived and stable; it has large gas giants nearby to protect it from debris impacts and help stabilise its orbit; it orbits within its star's habitable zone; has a large moon which stabilises its axial tilt; has a rocky composition; has a comfortable gravity which retains an atmosphere; has plate tectonics which regulate the climate and recycle vital elements; has a strong magnetic field which protects against radiation; has a chemical composition rich in essential elements, and so on [2]. Finding a planet with the same level of potential habitability would be challenging.

Even if a rare Earth was found, it would likely be so far away that transporting a large population to it could be simply infeasible [4] [5]. The galaxy is treacherously huge – traversing just 1% of its diameter at a speed of 0.1$c$ would take 10,000 years. The practicalities of transporting just a small number of people along with a livable biosphere across this distance presents enormous technological, ecological and sociological hurdles, due to the need for the passengers to inhabit the craft over this time period. The craft would effectively need to replicate a miniature Sun-Earth system, complete with sunlight, artificial gravity, agriculture, perfect recycling of water and vital elements, a stable ecology, manufacturing facilities, radiation shielding, and so on. A far more feasible technique could be to transport the passengers and associated biosphere in a low-metabolic stasis, if possible. Alternatively they could be transported as zygotes or even genetic code, with the colonists and biosphere automatically birthed and grown on arrival - however this would still leave the majority of the civilisation behind in the home system. The deciding factor for the migration strategy then could be the limits of biotechnology and the feasibility of reliably freezing into stasis and rethawing a population after many centuries, or alternatively, the realisation of much faster-than-light travel.

The migration strategy requires terraforming the destination

planet – the feasibility of which is uncertain. Planetary attributes such as gravity and day/night cycle would presumably not be feasible to change, whereas creating a comfortable Earth-like atmosphere may be. However even this may be infeasibly difficult – for example, Jokosky and Edwards' study concluded that creating an Earth-like atmosphere on Mars would be impossible using current technology [85].

The space habitation strategy would require the same level of technological, ecological and sociological mastery, but would be more feasible due to repairs and maintenance being able to utilise nearby bodies in the solar system.

Simply put, the safeguards would be more feasible than space habitation and migration by one and two orders of difficulty respectively: unmanned vehicles are easier to operate in space than populated craft, and interplanetary travel is easier than interstellar travel.

## 9.3 Risks

If a civilisation was able to transport a viable population to another star system, the objective would be for the settlers to develop into a full civilisation at the new system. However, the new civilisation would be cut off from its original one due to any two-way communications taking years to conduct, possibly over many generations. The settlers would also be under no obligation to continue contact. This presents the risk of ideological and technological drift and thus conflict. Even if all memory of the home system was somehow erased during the journey, the settlers would still present the risk of finding the home system eventually. A civilisation could send settlers out to a new system and after a few million years find themselves being visited and wiped out by a more technologically or militarily advanced civilisation desiring a rare Earth, akin to human history with native populations encountering more technologically advanced colonists. This risk would increase with time and number of fragmentations.

Conversely, fragmenting across many stellar systems hedges against the risk of being wiped out by a single astronomical event – however, astronomical events are predictable, whereas the long term development of civilisations are not. Fragmentation risk would thus outweigh astronomical risk. Therefore a civilisation considering the migration strategy would be incentivised to migrate in one large action as a complete civilisation, in order to avoid this fragmentation risk.

For the safeguard strategy, a bad actor could sabotage a safeguard which could cause catastrophe. However this risk of catastrophe is present in any technology of sufficient scale – a bad actor could equally sabotage the migration craft in transit, or the space habitats. Such risk is also present in our own civilisation in the shape of nuclear weapons, with mutually assured destruction disincentivising their use. A similar dynamic may hold for a civilisation that has built the safeguards – any bad actor which destroyed the system would only be ensuring their own destruction. Certain mechanisms could be employed to protect against any disruption, such as automation, the enforcement of a no-fly zone near the safeguards, and so on. The whole civilisation being reliant on the same system may incentivise closer social cohesion and stability.

On the technical side, the safeguard system presents the risk of catastrophe if there are any critical technical failures and, for example, the artificial sun ceases emitting sunlight – however this could be mitigated by redundancy and automated operation. Any repairs or maintenance can be carried out in the system – neighbouring celestial bodies hold large reserves of elements which could be used for repairs and maintenance over long timescales, with the space habitation strategy holding the same advantage. The space habitation strategy also benefits from having the population held in many separate bodies, providing redundancy. For the migration strategy, each craft transporting members of a civilisation would present the risk of the total loss of those passengers in transit – any critical problems which arise during the journey would be very difficult or impossible to correct, due to the extreme energy requirements of stopping at any intermediate systems for vital resources. However this risk could be reduced by migrating in many separate craft.

## 9.4 Benefits

With the safeguard system the reward is remaining in the civilisation's home system and on its home planet. With the migration strategy, the reward is a new planet where the quality of life would not be vastly superior to the rare Earth in the home system, and would likely be inferior. Given the time that it takes for intelligent life to evolve, Earth will be very comfortable to the civilisation, and would serve as the benchmark for the terraforming of any settled planets – finding a rare planet and terraforming it as well as possible would only produce planetary conditions roughly equal to Earth.

A civilisation choosing the safeguard strategy would be able to remain in the solar system without the difficulties of large-scale interstellar travel. It would also benefit from remaining on Earth with the historical-cultural value, infrastructure development, biosphere, and other assets which may be difficult or impossible to transport to an alternate stellar system.

If the civilisation was motivated towards the migration strategy because of damage to the Earth's biosphere, it would need to terraform any planets in the new system anyway – it would surely be more feasible simply to use those abilities to terraform Earth back to health.

One potential reward for successfully implementing the migration strategy would be a greater sociological strength for the civilisation. Being able to repeatedly terraform a planet in advance, load a biosphere and a large population into interstellar craft, and travel interstellar distances would necessitate a highly cohesive and organised civilisation, such that over time, factions of the civilisation which are less so would necessarily be filtered out.

Many large space habitats would be able to provide a total livable area 50 times Earth's land surface area, by utilising all of the asteroid belt's carbon to manufacture carbon nanotube for 461 km radius cylinders, per McKendree [86]. However unlike on Earth, these spinning habitats would place a limit on the number of habitable floors that can be stacked on this area, due to physiological constraints such as air pressure and gravity, which limits the top floor to 1,600 m in altitude, compared with the 5,000 m altitude at which humans can live on Earth [87]. Humans could also inhabit underground volumes below the Earth's surface – while the maximum ceiling height would be shorter than the habitats' at around 10 km due to the thickness of the Earth's crust, the inhabitants would have access to the open sky on the surface. The space habitats would be able to offer a third more total theoretical livable floor space than Earth, although this could increase with the invention of stronger materials or mining carbon from the Sun or gas giants. In contrast to the closed inner surface of the stations, some colonists may prefer Earth's solid ground, clear sky and open horizons as a permanent living environment.

The long-term survival of any civilisation, whichever strategy is being employed, would obviously be dependent on a near-zero probability of self-destruction per year – through war, biotechnology, ecological destruction, or similar. The civilisation could be a descendent of humanity's current civilisation, an alternate human civilisation, or an emergent non-human civilisation in the future. Humanity has made so much progress over just the past 100 years, with a promising current period of peace despite hugely increased population [88]. However, as summarised by Wilson, "the real problem of humanity is the following: We have Paleolithic emotions, medieval institutions, and Godlike technology" [89]. Since emotions are essential to the human experience, whether humanity can reduce its yearly self-destruction probability enough seems contingent on us improving our institutions to the same degree as our technology.

## 10 LIMITATIONS AND FUTURE STUDY

This study has focussed on the conceptual and resource aspects of the safeguards, and assumed that engineering solutions would be able to be maintained over very long timescales. Future studies could explore these aspects – more detailed exploration of the technical operation, and how might a civilisation ensure that the system works indefinitely using technologies with minimal long- term degradation. The sociological implications could also be explored. There may also be additional long-term natural threats which would require additional safeguards.

## 11 CONCLUSIONS

Seven long-term threats to the Earth's biosphere have been described and a set of safeguards investigated and conceptualised, in order to retain the habitability of the Earth long after the end of the Sun's natural life-supporting luminosity. The safeguards may be a more economically attractive, energy-inexpensive, and feasible long-term survival strategy as compared to alternative survival strategies.

If the safeguards were implemented and maintained, life on Earth would be able to continue as normal, enjoying comfortable security, geology and climate, and for 100 trillion years be able to look up at the night sky and see the stars – after which, natural star formation in the visible universe would end and the last stars would wink out. However, on Earth the Sun would still rise the next morning – with 9.0 million billion years of artificial sunlight remaining, the Earth's biosphere would still be very much in its infancy.

## REFERENCES

1. Andrew E. Snyder-Beattie, Anders Sandberg, K. Eric Drexler, and Michael B. Bonsall. "The timing of evolutionary transitions suggests intelligent life is rare" *Astrobiology*, **21**(3):265–278, 2021.
2. Peter Ward, Donald Brownlee, and Lawrence Krauss. "Rare Earth: Why complex life is uncommon in the universe". *Physics Today*, **53**, 09 2000.
3. K.-P. Schröoder and Robert Connon Smith. "Distant Future of the Sun and Earth revisited". *Monthly Notices of the Royal Astronomical Society*, **386**(1):155–163, 03 2008.
4. John Gertz. "There's no place like home (in our own solar system): Searching for ET near white dwarfs", *JBIS*, Vol 72, 2019, pp386-395.
5. Matthew T. Scoggins and David Kipping. "Lazarus stars: numerical investigations of stellar evolution with star-lifting as a life extension strategy". *Monthly Notices of the Royal Astronomical Society*, **523**(3):3251–3257, 06 2023.
6. NASA National Space Science Data Center. Planetary fact sheets, n.d.
7. Alessandro Morbidelli. *Origin and dynamical evolution of comets and their reservoirs,* 2005.
8. E. V. Pitjeva and N. P. Pitjev. "Masses of the main Asteroid Belt and the Kuiper Belt from the motions of planets and spacecraft". *Astronomy Letters*, **44**(8-9):554–566, 08 2018.
9. Bradley W. Carroll and Dale A. Ostlie. *An Introduction to Modern Astrophysics*. Benjamin Cummings, San Francisco, revised 2nd edition, 1995.
10. Larry Nittler, Nancy Chabot, Timothy Grove, and Patrick Peplowski. *The chemical composition of Mercury*. arXiv e-prints, 12 2017.
11. David H. Hathaway, R. M. Wilson, and E. J. Reichmann. "The Sun's shallow meridional circulation". *The Astrophysical Journal,* **644**(1):598–603, 2006.
12. Richard G. French, Matthew M. Hedman, Philip D. Nicholson, Pierre-Yves Longaretti, and Colleen A. McGhee-French. "The Uranus system from occultation observations (1977-2006): Rings, pole direction, gravity field, and masses of Cressida, Cordelia, and Ophelia". *Icarus*, **411**:115957, 03 2024.
13. M. E. Caplan. "Black dwarf supernova in the far future". *Monthly Notices of the Royal Astronomical Society*, **497**(4):4357–4362, August 2020.
14. Brian Lacki. *Ground to dust: Collisional cascades and the fate of Kardashev II Megaswarms.* arXiv e-prints, 04 2025.
15. Abraham Loeb. "Interstellar objects from broken Dyson Spheres." *Research Notes of the AAS*, **7**(3):43, 03 2023.
16. Alexander A. Svoronos. "The star tug: An active stellar engine capable of accelerating a star to relativistic velocities". *Acta Astronautica*, **176:**306–312, 2020.
17. B. R. Finney and E. M. Jones. *Solar System Industrialization: Implications for Interstellar Migrations*. University of California Press, Berkeley, CA, 1985.
18. Stuart Armstrong and Anders Sandberg. "Eternity in six hours: Intergalactic spreading of intelligent life and sharpening the Frmi Paradox". *Acta Astronautica*, **89**:1–13, 2013.
19. Bill Paxton, Lars Bildsten, Aaron Dotter, Falk Herwig, Pierre Lesaffre, and Frank Timmes. "Modules for experiments in stellar astrophysics (mesa)". *Astrophysical Journal Supplement*, **192**(1):3, 01 2011.
20. James F. Kasting, Daniel P. Whitmire, and Ray T. Reynolds. "Habitable zones around main sequence stars". *Icarus*, **101**(1):108–128, 01 1993.
21. István Szapudi. "Solar radiation management with a tethered sun shield". *Proceedings of the National Academy of Sciences*, **120**(32):e2307434120, 2023.
22. Dimitri Veras and Aline A. Vidotto. "Planetary magnetosphere evolution around post-main-sequence stars". *Monthly Notices of the Royal Astronomical Society*, **506**(2):1697–1703, 06 2021.
23. Richard Dendy. *Plasma Physics: An Introductory Course*. Cambridge University Press, 1995.
24. J.-H. Shue, P. Song, C. T. Russell, J. T. Steinberg, J. K. Chao, G. Zastenker, et al. *Solar wind carbon, nitrogen, and oxygen abundances and their dependencies on solar magnetic structures*. arXiv e-prints, 1986.
25. Alexey S. Kurlov and Aleksandr I. Gusev. *Tungsten Carbides: Structure, Properties and Application in Hardmetals*. Springer Science & Business Media, 2013.
26. M.F. Yu, O. Lourie, M.J. Dyer, K. Moloni, T.F. Kelly, and R.S. Ruoff. "Strength and breaking mechanism of multiwalled carbon nanotubes under tensile load". *Science*, **287**(5453):637–640, January 2000.

27. S. Kotrechko, I. Mikhailovskij, T. Mazilova, et al. "Mechanical properties of carbyne: experiment and simulations". *Nanoscale Research Letters*, **10**:24, 2015.

28. Joseph Minow, William Blackwell, Linda Neergaard, Steven Evans, Donna Hardage, and Jerry Owens. "Charged particle environment for ngst: L2 plasma environment statistics". *Proceedings of SPIE – The International Society for Optical Engineering*, pages 942-953, 07 2000.

29. G. Gloeckler, F. M. Ipavich, D. C. Hamilton, B. Wilken, W. Stüdemann, G. Kremser, and D. Hovestadt. "Solar wind carbon, nitrogen and oxygen abundances measured in the earth's magnetosheath with ampte/cce" *Geophysical Research Letters*, **13**(8):793-796, 1986.

30. NASA. *Webb's Sunshield*, 2025.

31. Kan Yang, Michael Choi, Eric Mentzell, and Andrew Concha. "JWST sunshield thermal performance evaluation". Technical Report AIAA 2010-6048, NASA Goddard Space Flight Center, 2010.

32. G. Jeffrey Taylor and Mark A. Wieczorek. "Lunar bulk chemical composition: a post-gravity recovery and interior laboratory reassessment". *Philosophical Transactions. Series A, Mathematical, physical, and engineering sciences,* **372**(2024):20130242, September 2014.

33. J. H. Davies and D. R. Davies. *Earth's surface heat flux. Solid Earth*, **1**(1):5–24, 2010.

34. D. C. Ferguson, M. E. Palaszewski, and R. J. Litchford. "Atmospheric mining in the outer solar system: Resource capturing, exploration, and exploitation". Technical Report NASA/TM–2015-218831, NASA Glenn Research Center, 2015.

35. Matthew E. Caplan. "Stellar engines: Design considerations for maximizing acceleration". *Acta Astronautica,* **165**:96–104, 2019.

36. Bradley W. Carroll and Dale A. Ostlie. *An Introduction to Modern Stellar Astrophysics*. Addison Wesley, San Francisco, 2007.

37. Royal Belgian Institute for Space Aeronomy. *Jupiter: Gas giant planetary facts*, 2025.

38. Michele K. Dougherty, Larry W. Esposito, and Stamatios M. Krimigis. "Saturn after Cassini-Huygens". In *Saturn from Cassini-Huygens*, pages 745-761. Springer, 2009.

39. B. Conrath, D. Gautier, R. Hanel, G. Lindal, and A. Marten. "The Helium abundance of Uranus from Voyager measurements". *Journal of Geophysical Research*, **92**(A13):15003-15010, 12 1987.

40. W. B. Hubbard. "Neptune's deep chemistry". *Science*, **275**(5304):127-1280, 1997.

41. Ravit Helled and Saburo Howard. *Giant planet interiors and atmospheres*, 2024.

42. Pejvak Javaheri, Hanno Rein, and Dan Tamayo. "Whfast512: A symplectic n-body integrator for planetary systems optimized with avx512 instruc tions". *The Open Journal of Astrophysics*, **6**:29, 07 2023.

43. A. Higuchi and S. Ida. "Temporary capture of asteroids by an eccentric planet". *The Astronomical Journal*, **153**(4):155, 04 2017.

44. John F. Ready. *Industrial Applications of Lasers*. Academic Press, San Diego, 2 edition, 1997.

45. Lucas Norder, Shunyu Yin, Matthijs H. J. de Jong, Francesco Stallone, Hande Aydogmus, Paolo M. Sberna, Miguel A. Bessa, and Richard A. Norte. "Pentagonal photonic crystal mirrors: scalable lightsails with enhanced acceleration via neural topology optimization". *Nature Communications*, **16**(1):2753, March 2025.

46. Attay Kovetz, Ofer Yaron, and Dina Prialnik. "A new, efficient stellar evolution code for calculating complete evolutionary tracks". *Monthly Notices of the Royal Astronomical Society*, **395**(4):1857-1874, 05 2009.

47. D. G. Korycansky, Gregory Laughlin, and Fred C. Adams. "Astronomical engineering: A strategy for modifying planetary orbits". *Astrophysics and Space Science*, **275**(4):349-366, 2001.

48. A. Hyder, C. Li, N. Chanover, and G. Bjoraker. "A supersolar oxygen abundance supported by hydrodynamic modelling of Jupiter's atmosphere". *Nat Astron*, **9**(2):211–220, 2025.

49. Muhammad Amir, Muhammad Saqib, Shoukat Hayat, Abdul Rauf, Saleh N. Alanazi, Muhammad Rizwan, Adeel Raza, Ammar Mehmood, and Younas Khan. "Lightning for energy and material uses: A structured review". *Frontiers in Energy Research*, **8**:183, 2020.

50. Donald L. Turcotte and Gerald Schubert. *Geodynamics*. Cambridge University Press, Cambridge, 2002.

51. B. A. Buffett. "Taking Earth's temperature". *Science*, **315**(5820):1801-1802, 2007.

52. Christopher J. Spencer. "Biogeodynamics: Coupled evolution of the biosphere, atmosphere, and lithosphere". *Geology*, **50**(8):867-868, 08 2022.

53. James W. Valentine. "Plate tectonics and the biosphere". *Paleontology*, pages 622-627. Springer, Berlin, Heidelberg, 1979.

54. Robert J. Stern and Taras V. Gerya. "The importance of continents, oceans and plate tectonics for the evolution of complex life: implications for finding extraterrestrial civilizations". *Scientific Reports*, **14**:8552, 2024.

55. Qiuming Cheng. "Extrapolations of secular trends in magmatic intensity and mantle cooling: Implications for future evolution of plate tectonics". *Gondwana Research*, 63:268-273, 2018.

56. United States Geological Survey. "Why are we having so many earthquakes? has naturally occurring earthquake activity been increasing?", 2025. https://www.usgs.gov/faqs/why-are-we-having-so-many-or-so-few-earthquakes-has-naturally-occurring-earthquake-activity [accessed 8 June 2026]'

57. David J. Stevenson. "Planetary science: Mission to Earth's core—a modest proposal". *Nature*, **423**(6937):239-240, 05 2003.

58. Sawsan Omira and Abdel Mourad. "Future of antimatter production, storage, control, and annihilation applications in propulsion technologies". *International Journal of Thermofluids*, **25**:101012, 12 2024.

59. PeriodicTable.org. *Iron thermal expansion coefficient*, 2024.

60. William M. White and Emily M. Klein. "Elemental fluxes and the composition of the oceanic crust". *Earth and Planetary Science Letters*, 371-372:1-14, 2014.

61. L. M. Kistler, A. B. Galvin, M. A. Popecki, K. D. C. Simunac, C. Farrugia, E. Moebius, M. A. Lee, L. M. Blush, P. Bochsler, P. Wurz, B. Klecker, R. F. Wimmer-Schweingruber, A. Opitz, J.-A. Sauvaud, B. Thompson, and C. T. Russell. "Escape of O+ through the distant tail plasma sheet". *Geophysical Research Letters*, **37**(21), 2010.

62. Peter H. Gleick. "Water resources" in *Encyclopedia of Climate and Weather*, pages 817-823. Oxford University Press, New York, 1996.

63. S. Alan Stern. "The evolution of comets in the Oort Cloud and Kuiper Belt". *Nature*, **424**(6949):639–642, 2003.

64. Olga P. Popova, Peter Jenniskens, Vacheslav Emel'yanenko, Anna Kartashova, Eugeny Biryukov, Sergey Khaibrakhmanov, Valery Shuvalov, Yurij Rybnov, Alexandr Dudorov, Victor I. Grokhovsky, et al. "Chelyabinsk airburst, damage assessment, meteorite recovery, and characterization". *Science*, **342**(6162):1069-1073, 2013.

65. James G. Williams and Doris H. Boggs. "Secular tidal changes in lunar orbit and Earth rotation". *Celestial Mechanics and Dynamical Astronomy*, **126**(1):89-129, 2016.

66. F. R. Stephenson, L. V. Morrison, and C. Y. Hohenkerk. "Measurement of the Earth's rotation: 720 BC to AD 2015". *Proceedings of the Royal Society A: Mathematical, Physical and Engineering Sciences*, **472**(2196):20160404, 2016.

67. NASA Science. *Tidal Locking*, 2025.

68. O. Neron de Surgy and J. Laskar. "On the long term evolution of the spin of the Earth". *Astronomy and Astrophysics*, **318**:975-989, 02 1997.

69. Jack J. Lissauer, Jason W. Barnes, and John E. Chambers. "Obliquity variations of a moonless Earth". *Icarus*, **217**(1):77–87, 2012.

70. Arnold Hanslmeier. "Habitability and cosmic catastrophes". In *Habitability and Cosmic Catastrophes*, 2008.

71. NASA Earth Observations (NEO). Solar Insolation (1 month), 2025.

72. Center for Near Earth Object Studies. *Discovery Statistics*, 2025.

73. Karen Meech, Robert Weryk, Marco Micheli, Jan Kleyna, Olivier Hainaut, Robert Jedicke, Richard Wainscoat, Kenneth Chambers, Jacqueline Keane, Andreea Petric, Larry Denneau, Eugene Magnier, Travis Berger, Mark Hu- ber, Heather Flewelling, C. Waters, Eva Schunova-Lilly, and Serge Chastel. "A brief visit from a red and extremely elongated interstellar asteroid". *Nature*, **552**, 12 2017.

74. David Morrison, Clark R. Chapman, Duncan Steel, and Richard P. Binzel. "Impacts and the public: Communicating the nature of the impact hazard". In *Mitigation of Hazardous Comets and Asteroids*. Cambridge University Press, Cambridge, 2004.

75. Andrew F. Cheng, Elisabetta Dotto, Olivier S. Barnouin, Saverio Cambioni, Andrew S. Rivkin, Cristina A. Thomas, Nancy L. Chabot, R. Terik Daly, Carolyn M. Ernst, Sabina D. Raducan, et al. "Momentum transfer from the Dart mission kinetic impact on asteroid Dimorphos". *Nature*,

616(7956):447-450, 2023.

76. R. Terik Daly, Olivier S. Barnouin, Nancy L. Chabot, Carolyn M. Ernst, Andrew F. Cheng, Andrew S. Rivkin, Alexander G. Hayes, Cristina A. Thomas, Elisabetta Dotto, Zachary J. Fletcher, et al. "Successful kinetic impact into an asteroid for planetary defence". *Nature*, **616**(7956):438-442, 2023.

77. Dana G. Andrews, K. D. Bonner, A. W. Butterworth, H. R. Calvert, B. R. H. Dagang, Simon Dimond, L. G. Eckenroth, J. M. Erickson, B. A. Gilbertson, N. R. Gompertz, O. J. Igbinosun, T. J. Ip, B. H. Khan, and S. L. Marquez. "Defining a successful commercial asteroid mining program." *Acta Astronautica,* **108**:1-118, 2015.

78. J. Laskar and M. Gastineau. "Existence of collisional trajectories of Mercury, Mars and Venus with the Earth". *Nature*, **459**(7248):817-819, 2009.

79. Milan M. Církovićand Branislav Vukotić. "Long-term prospects: Mitigation of supernova and gamma-ray burst threat to intelligent beings". *Acta Astronautica*, **129**:438-446, 2016.

80. Sean N. Raymond, Nathan A. Kaib, Franck Selsis, and Herve Bouy. "Future trajectories of the solar system: dynamical simulations of stellar encounters within 100 au", 2023. https://arxiv.org/abs/2311.12171 [accessed 8 June 2026]

81. T. J. Cox and Abraham Loeb. "The collision between the Milky Way and Andromeda". *Monthly Notices of the Royal Astronomical Society*, 386(1):461– 474, 03 2008.

82. Clément Vidal. *The spider stellar engine: a fully steerable extraterrestrial design?* arXiv e-prints, 11 2024.

83. Jonathan H. Jiang, Fuyang Feng, Philip E. Rosen, Kristen A. Fahy, Prithwis Das, Piotr Obacz, Antong Zhang, and Zong-Hong Zhu. "Avoiding the great filter: Predicting the timeline for humanity to reach Kardashev type I civilization. *Galaxies*, **10**(3):68, 2022.

84. Gerard K. O'Neill. The colonization of space. *Physics Today*, **27**(9):32–40, sep 1974. Bibcode: 1974PhT. 27i..32O.

85. B. M. Jakosky and C. S. Edwards. Inventory of $CO_2$ available for terraforming Mars". *Nature Astronomy*, 2:634-639, 2018.

86. Tom McKendree. "Appropriately ambitious aerospace goals: Keynote address" to the Turning Goals Into reality conference, May 2000.

87. David W. Jensen. *Design limits on large space stations*, 2024.

88. Bruno Tertrais." The demise of Ares: The end of war as we know it? *The Washington Quarterly*, **35**(3):7–22, 2012.

89. E. O. Wilson. Debate at the Harvard museum of Natural History, 2009.

90. Paul Birch, "Orbital Ring Systems and Jacob's Ladders - I", Journal of the British Interplanetary Society, Vol. 35, 1982, pp. 475– 497.

Version 2 (9th August 2026) – changed sunshade composition from aluminium to alumina fabric; added Birch orbital ring for Jupiter mass lifting.